# Control and synchronization of capillary flows in stepped microchannels

H. Desu, N. S. Satpathi, L. Malik, and A. K. Sen

Micro Nano Bio Fluidics Unit, Department of Mechanical Engineering, Indian Institute of Technology Madras, Chennai – 600036, Tamil Nadu, India.

Corresponding Author: ashis@iitm.ac.in

## ABSTRACT

Capillary-driven transport offers a simple, self-sustained alternative to externally pumped microfluidic systems, yet achieving precise control of such flows remains challenging. We experimentally and theoretically investigate capillary flow in rectangular microchannels when a liquid meniscus encounters a geometric step with varying channel width and height. Depending on the contact angle and step dimensions, the meniscus either pins or advances, defining distinct flow and no-flow regimes. Introducing a lateral offset between channel sidewalls provides an additional geometric control that enables reversible switching from a pinned to a flowing state, even for liquids with relatively high contact angles. We rationalize these transitions using numerical simulations and an energy-based scaling model that captures the balance between surface tension and Laplace pressure at the step. Finally, we demonstrate synchronization of capillary fronts in parallel microchannels by combining stepped and offset geometries. These results establish a simple, geometry-mediated mechanism for passive control and synchronization of capillary flows, expanding the design space for microfluidic systems.

## I. INTRODUCTION

The precise control and manipulation of microliter-scale liquids in confined geometries remain central challenges in microfluidics [1]. A fundamental understanding of capillary transport in microchannels is crucial for enabling reliable operation across diverse applications, from biochemical analysis [2] to electronics cooling [3]. Despite advances in microfluidic technology, most existing systems still rely on external pumping mechanisms that impose stringent requirements on precision, stability, and cost [4, 5]. These limitations motivate the development of autonomous, geometry-guided flow-control strategies that exploit interfacial forces rather than external actuation.

Capillary-driven flow presents a promising route to achieve self-sustained fluid transport in microchannels. Such flows arise from a balance between Laplace pressure, surface tension, and viscous resistance, allowing passive operation without external inputs. However, realizing controllable capillary transport remains challenging. Active microvalves, based on pneumatic [7], electromagnetic [8, 9], electrostatic [10], or photothermal [11] actuation, offer dynamic control but add mechanical and fabrication complexity. In contrast, passive microvalves [6] employ geometric features to manipulate the interfacial curvature and Laplace pressure, enabling flow modulation through intrinsic physical mechanisms.

Several geometric strategies have been explored to regulate capillary flow. Patterned wettability [12] and phase guides [13–15] can locally modify interfacial curvature to slow, stop, or redirect flow. Spacer-based synchronization of parallel capillary channels has been demonstrated, where meniscus pinning at obstacles equalizes flow velocities [16]. Step- or expansion-based capillary stop valves have been used to trigger flow control [17], with analytical and numerical models developed to predict burst pressures and meniscus stability in divergent geometries [18–22]. Energy-based approaches have also been applied to determine spontaneous flow conditions across backward-facing steps [23]. More recently, virtual air-liquid interfaces have been used to actuate flow through capillary gating [24], while complex stop-valve circuits have enabled autonomous sequencing for biochemical assays [25].

While these studies have advanced capillary valve design, a comprehensive understanding of how geometric discontinuities and liquid properties jointly govern meniscus stability is still lacking. In particular, the interplay between step aspect ratio, contact angle, and local curvature that determines whether a meniscus advances or halts has not been systematically quantified. Here, we experimentally and theoretically investigate capillary flow dynamics in rectangular microchannels when the advancing meniscus encounters a geometric step composed of width and height discontinuities. We construct a regime map that delineates flow and no-flow states as functions of step geometry and contact angle. Furthermore, by introducing an offset between the channel side walls, we reveal an additional control parameter that can trigger a transition from a pinned to a flowing state, even for liquids with high contact angles and large steps.

Our analysis combines experiments, numerical simulations, and energy-based scaling to show that the flow/no-flow boundary corresponds to the sign of the net surface energy change, $\Delta E$, across the step. Flow occurs when $\Delta E < 0$, i.e., when the Laplace pressure remains favorable to meniscus advancement. The offset step geometry stabilizes a concave meniscus that maintains a positive Laplace pressure, thereby expanding the operational envelope for spontaneous capillary transport. Finally, we demonstrate synchronized flow in parallel microchannels by integrating stepped and offset-stepped geometries, offering a generalizable framework for passive flow regulation.

Beyond microfluidic implementation, this study provides a physics-based description of geometry-mediated capillary stability, establishing predictive relations between interfacial curvature, contact angle, and confinement geometry. The presented regime map and scaling framework offer new insights into metastable meniscus dynamics and introduce a minimal, energy-driven approach to controlled capillary flow in confined systems.

## III. EXPERIMENTAL

### A. Device geometry and fabrication

The schematics of the two capillary flow devices used in this study are depicted in Fig. 1a. Both feature an inlet channel of width $w_1$ and height $h_1$, and an outlet channel of width $w_2$ and height $h_2$, connected by a step geometry. The step valve (SV) design has symmetric steps along the channel width. In contrast, the offset-step valve (OSV) introduces asymmetry due to a difference in the lengths of the microchannel side walls, creating an offset denoted by $g$. The ranges of dimensions of the microchannel studied in the present work are as follows: $h_1 = 0.1 - 0.4\ mm$, $w_1 = 0.1 - 0.4\ mm$, $h_2 = 0.2 - 0.8\ mm$, $w_2 = 0.6 - 1.6\ mm$, and $G = 0 - 0.8\ mm$. The microchannels used in the capillary flow experiments were fabricated on polymethyl methacrylate (PMMA) substrates via CNC micromilling, followed by solvent-assisted thermal bonding [26]. The channel structures, including the step features, were milled into the bottom substrate, while the inlet and outlet ports were machined into the top cover plate using a CNC micromilling machine (Mini-Mill/3, Minitech Machinery, USA). For bonding, the mating surfaces of the substrates were exposed to chloroform vapor for 30 s to induce surface softening, after which the substrates were aligned and bonded under a pressure of 10 bar at a temperature of 100 °C. Precise alignment of the channel structures with the inlet and outlet ports was achieved using dedicated alignment holes fabricated in both substrates and corresponding alignment pins. Post-bonding, the assembled devices were conditioned at ambient temperature for 24 h to allow relaxation of any residual thermal or mechanical stresses before experimental use.

### B. Materials and methods

Water-ethanol mixtures with ethanol concentrations ranging from 0% to 60% v/v were used as working fluids in the experiments. The corresponding physical properties of these mixtures are summarized in Table 1. The static contact angle of each fluid on the PMMA surface was measured using a goniometer (DSA25B, KRÜSS, Germany), yielding values in the range of 36° to 72°. The dynamic viscosity of the mixtures was determined using a rotational viscometer (MCR series, Anton Paar, Germany), with measured values ranging from $1.003 \times 10^{-3}$ to $3.070 \times 10^{-3}$ Pa-s. Surface tension measurements were performed using a tensiometer (Sigma 702, Biolin Scientific), and the values were found to vary between 72.88 and 28.95 mN/m, depending on the ethanol concentration.

A schematic of the experimental setup is shown in Fig. 1b. In each experiment, a fixed liquid volume of 10 µL, approximately equal to the internal volume of the microchannel, was dispensed at the device inlet. To maintain purely capillary-driven flow, the dispensed volume was minimized, and the inlet reservoir diameter was kept relatively large (~4 mm), thereby reducing any significant hydrostatic pressure at the channel entrance. The top-view dynamics of the advancing meniscus were recorded using an inverted microscope (IX73, Olympus, Japan) coupled with a high-speed camera (V2670, Phantom, USA). For front-view visualization of the meniscus motion, a separate high-speed imaging system (Fastcam Mini AX200, Photron, Japan) equipped with a 12X zoom lens (Nikkor AF-S, Nikon, Japan) was employed. Additionally, the synchronization of flow between parallel microchannels was captured using a USB digital microscope (Dinolite 3.0, Taiwan).

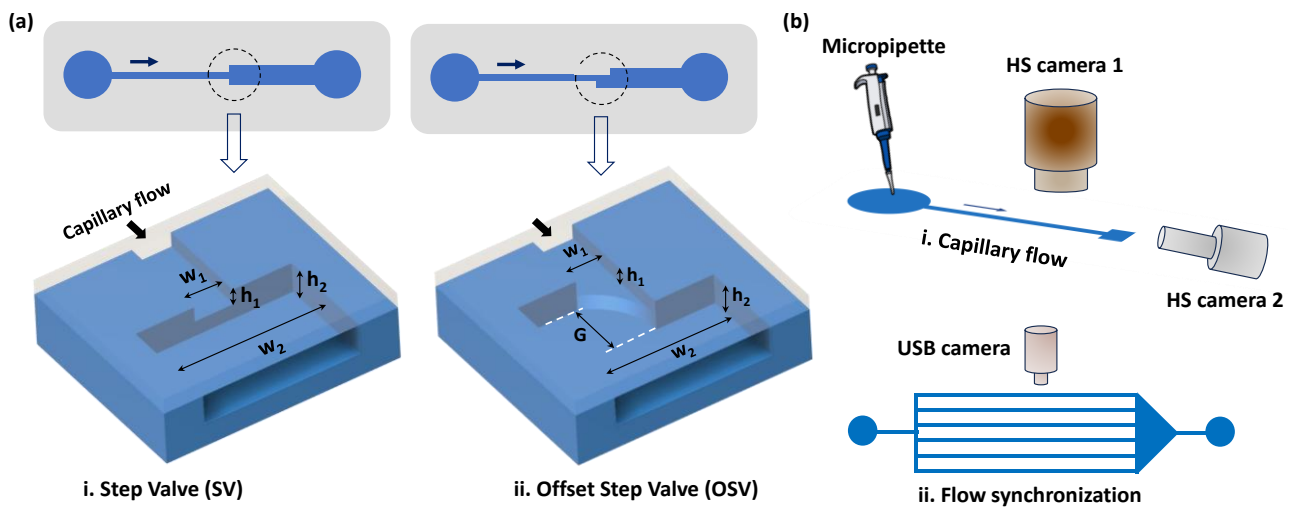

**Fig. 1:** (a) Schematic of the capillary flow devices with stepped microchannels, (i) Step valve (SV) with symmetric steps in the width direction, (ii) Offset-step valve (OSV) with asymmetric steps, due to a difference in the lengths of the channel side walls. Both valves have a step in the height direction. (b) Schematic of the experimental setup used for the capillary flow studies: (i) single channel with SV or OSV, (ii) Flow synchronization in multiple parallel channels with SV or SV+OSV.

**Table 1:** The properties of the water/ethanol mixture. The uncertainty in the measurement of contact angle, viscosity, and Surface tension is ±2°, ±0.001 Pa-s, and ±1 N/m, respectively.

| Ethanol/Water %v/v | Contact angle (°) | Dynamic viscosity (Pa-s) | Surface tension (N/m) |
|---|---|---|---|
| 0 | 72 | 1.003 | 72.88 |
| 10 | 67 | 1.288 | 53.43 |
| 20 | 63 | 1.810 | 43.71 |
| 30 | 57 | 2.313 | 37.16 |
| 40 | 53 | 2.616 | 33.88 |
| 50 | 42 | 3.079 | 31.36 |
| 60 | 36 | 3.070 | 28.95 |

## III. NUMERICAL MODEL

Three-dimensional (3D) numerical simulations are carried out using COMSOL Multiphysics to understand the meniscus dynamics and dynamic variation in the pressure jump across the meniscus while passing through the step, both in the case of the step valve and the offset-step valve. The capillary fluid flow is described by the conservation of mass and momentum via the Navier-Stokes equation with the surface tension force included as a body force to account for the capillary effects [27],

$$\rho \frac{\partial \vec{u}}{\partial t} + \rho(\vec{u} \cdot \nabla)\vec{u} = \nabla \cdot [-pI + \mu(\nabla \vec{u} + (\nabla \vec{u})^T)] + \vec{F}_{st} \qquad (1)$$

$$\rho \nabla \cdot \vec{u} = 0$$

where $\vec{u}$ is flow velocity, $\rho$ is fluid density, $p$ is fluid pressure, $I$ represent the identity matrix, $\mu$ is dynamic viscosity and $F_{st}$ represents surface tension force, whose expression is presented later.

The two-phase flows describing the capillary liquid and the air phases are modelled using the phase field model present in COMSOL Multiphysics. In the phase field model, two-phase systems are governed by the Cahn-Hilliard equation [28], given as follows,

$$\rho \frac{\partial \phi}{\partial t} + \vec{u} \cdot \nabla \phi = \nabla \cdot \left(\frac{\gamma \lambda}{\varepsilon_{pf}^2}\right) \nabla \psi \qquad (2)$$

where the chemical potential,

$$\psi = -\nabla \cdot (\varepsilon_{pf}^2 \nabla \phi) + (\phi^2 - 1)\phi + \frac{\varepsilon_{pf}^2}{\lambda}\left(\frac{\partial f}{\partial \phi}\right)$$

Here, $\phi$ is the dimensionless phase field parameter representing the volume fractions of the phases, which can have a numerical value of 0 or 1, and $\gamma$ is the mobility parameter that controls the relaxation time associated with the minimization of free energy represented by the terms in the right-hand side of equation 2, $\lambda$ (N) is the mixing energy density, and $\varepsilon_{pf}$ (m) is a capillary width that scales with the thickness of the interface. Here, $f$ is an energy density term originating from other physical processes such as elastic energy, temperature, or concentration fields and $f = 0$ in the present case. The two parameters ($\lambda$ and $\varepsilon_{pf}$) are related to the surface tension coefficient, $\sigma$ (N/m), through the equation,

$$\sigma = \frac{2\sqrt{2}}{3}\frac{\lambda}{\varepsilon_{pf}} \qquad (3)$$

The volume fractions of fluids at the interface are defined as

$$V_{\text{liq}} = \frac{(1-\phi)}{2}, V_{\text{gas}} = \frac{(1+\phi)}{2} \qquad (4)$$

Total density and the viscosity of the mixture at the interface are defined as

$$\rho = \rho_{\text{liq}} + (\rho_{\text{gas}} - \rho_{\text{liq}})V_{\text{gas}} \qquad (5)$$

$$\mu = \mu_{\text{liq}} + (\mu_{\text{gas}} - \mu_{\text{liq}})V_{\text{gas}}$$

The wetted wall condition is given by

$$n \cdot \left(\frac{\gamma \lambda}{\varepsilon_{pf}^2}\right) \nabla \psi = 0 \qquad (6)$$

$$n \cdot \varepsilon_{pf}^2 \nabla \phi = \varepsilon_{pf}^2 \cos\theta_w |\nabla \phi|$$

where $n$ is the unit vector normal to the wall surface and $\theta_w$ is the wall contact angle. This condition is typically used to implement a general wetting boundary condition that allows the fluid-fluid interface to move along the wall while satisfying specific contact angle conditions. The surface tension force in equation 1 is defined as

$$\vec{F}_{st} = K \, \nabla \phi, \, K = \frac{\lambda}{\varepsilon_{pf}^2} \, \phi \tag{7}$$

In COMSOL Multiphysics, the above equations are simultaneously solved to determine the meniscus dynamics. The boundary conditions and initial conditions used in the numerical simulations are as follows: the inlet reservoir connected to the capillary microchannel is filled with a liquid having a fixed set of properties, while the microchannel initially contains air. The initial velocity throughout the system is set to zero. At the inlet, a pressure boundary condition is applied with the pressure set to zero. As liquid begins to flow from the inlet reservoir into the microchannel, the volume fraction of water at the inlet is set to one. The flow inside the microchannel occurs solely due to the capillary forces. The outlet is assigned a pressure outlet boundary condition with the pressure set to zero. The four walls of the microchannel are subjected to a no-slip boundary condition. Additionally, a wetted wall boundary condition, as mentioned in equation 6, is applied to the channel walls, and the contact angle is specified to account for the interaction between the fluid and the channel surface.

In order to ensure that the results are mesh independent, a mesh convergence study is performed. The predefined mesh settings in COMSOL Multiphysics are used with the following maximum element sizes: coarser mesh (0.102 mm), coarse mesh (0.0788 mm), normal mesh (0.0528 mm), fine mesh (0.0418 mm) and finer mesh (0.0292 mm). The simulation results for the different mesh sizes are compared by studying the variation of meniscus position along the incoming channel before the step with time, as shown in Fig. 2a. A considerable difference in the results is observed as we go from the coarser mesh to the fine mesh, and the improvements from fine to finer mesh is found to be insignificant. Therefore, for the simulation study presented here, we proceed with a fine mesh with 734,423 elements and 133,449 vertices in the computational domain to preserve numerical accuracy while avoiding excessive computational time. A comparison of the simulation results and experimental data in terms of the meniscus displacement versus time is shown in Fig. 2b, which shows a good agreement.

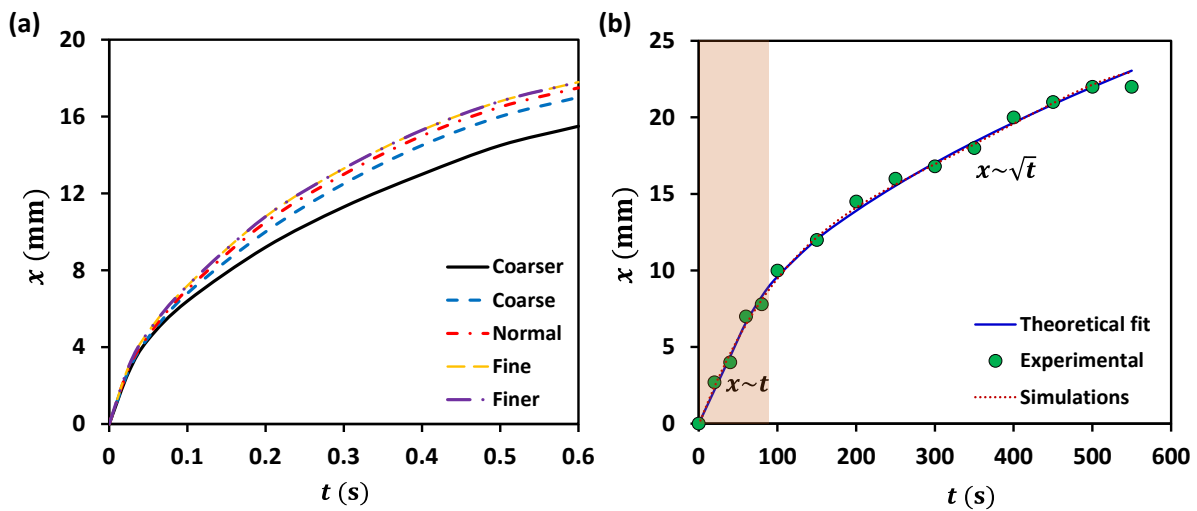

**Fig. 2:** (a) Comparison of simulation results with different mesh sizes by studying the variation of meniscus position with time, incoming channel with $w_1 = 0.4 \, mm$, and $h_1 = 0.4 \, mm$, $\theta_s = 53°$, (b) Comparison of simulation and experimental data in terms of the variation of meniscus position with time, with $w_1 = 0.4 \, mm$, $h_1 = 0.4 \, mm$, $w_2 = 1.6 \, mm$, and $h_2 = 0.5 \, mm$, and $\theta_s = 42°$.

## III. RESULTS & DISCUSSION

### A. Meniscus dynamics at the geometric step

We present the experimental images of the capillary flow meniscus at the geometric step both in the case of the step valve (SV) and the offset step valve (OSV), with water + ethanol as the working liquid, as shown in Fig. 3. In the case of a SV with $w_1 = 0.4 \, mm$, $h_1 = 0.4 \, mm$, $w_2 = 1.6 \, mm$, and $h_2 =$ $0.5\ mm$, and $\theta_s = 42°$, we find the meniscus flows past the step, indicating a flow regime (see Fig. 3a and Movie S1). However, for the same channel geometry, when a working liquid with $\theta_s = 57°$ is used, the meniscus fails to advance past the step, indicating a no-flow regime (see Fig. 3b and Movie S2). In the case of OSV, with an offset $G = 0.8\ mm$, for $w_1 = 0.4\ mm$, $h_1 = 0.4\ mm$, $w_2 = 1.6\ mm$, and $h_2 = 0.5\ mm$, and $\theta_s = 57°$, and we observe that the meniscus can flow past the step (see Fig. 3c and Movie S3). This highlights that for the same dimensions of the incoming and expanded channels and liquid contact angles, the introduction of an offset can lead to a transition from the no-flow regime to the flow regime. However, for a much higher liquid static contact angle of $\theta_s = 67°$, a no-flow regime is observed even in the case of OSV (see Fig. 3d and Movie S4).

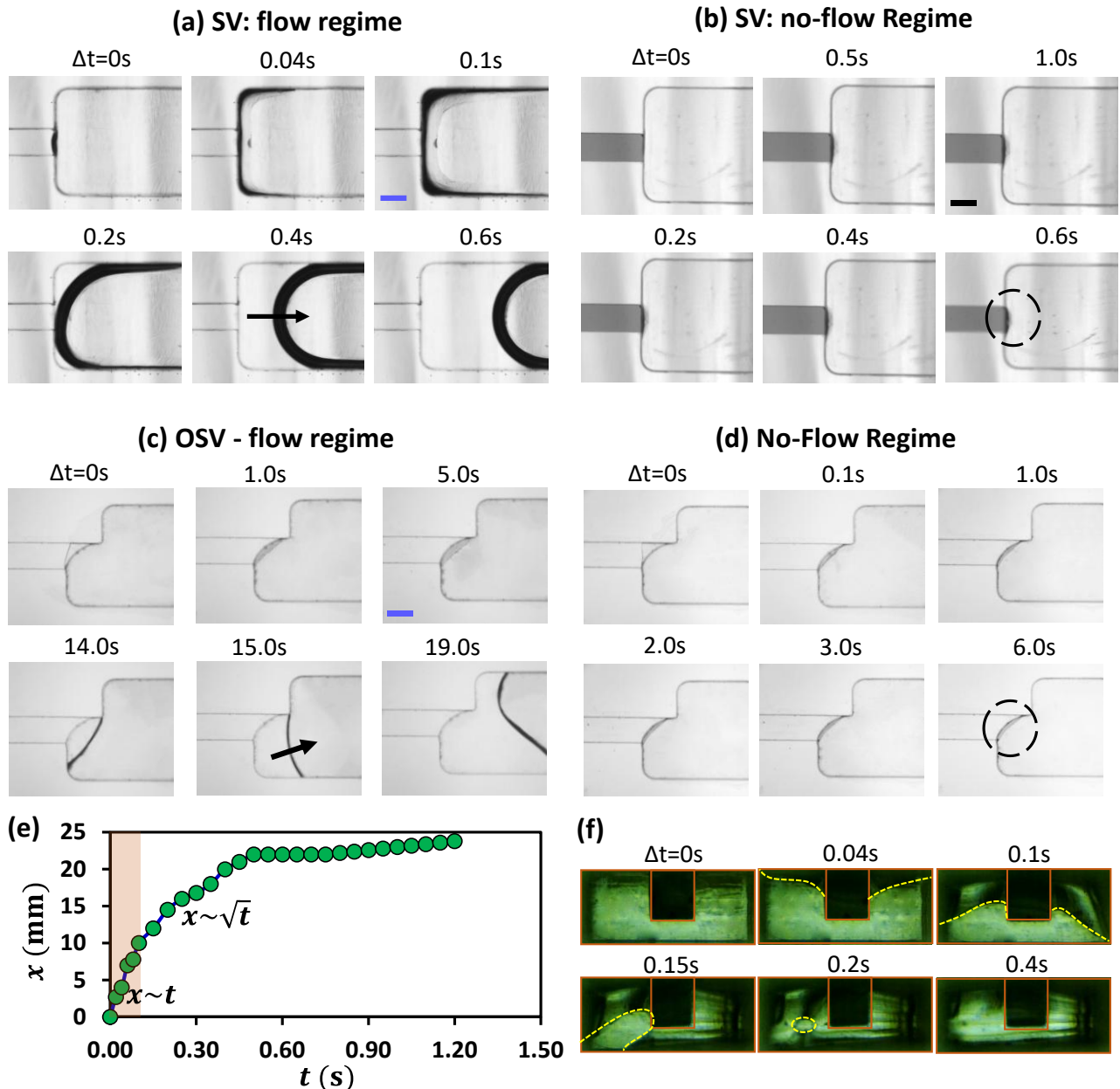

**Fig. 3:** Experimental top-view images of the capillary flow meniscus at the step valve (SV) and the offset step valve (OSV), with water + ethanol as the working liquid: (a) SV: $w_1 = 0.4\ mm$, $h_1 = 0.4\ mm$, $w_2 = 1.6\ mm$, and $h_2 = 0.5\ mm$, and $\theta_s = 42°$. (b) SV: $w_1 = 0.4\ mm$, $h_1 = 0.4\ mm$, $w_2 = 1.6\ mm$, $h_2 = 0.5\ mm$, and $\theta_s = 57°$. (c) OSV: $w_1 = 0.4\ mm$, $h_1 = 0.4\ mm$, $w_2 = 1.6\ mm$, $h_2 = 0.5\ mm$, $\theta_s = 57°$ and $G = 0.6\ mm$. (d) OSV: $w_1 = 0.4\ mm$, $h_1 = 0.4\ mm$, $w_2 = 1.6\ mm$, $h_2 = 0.5\ mm$, $\theta_s = 67°$ and $G = 0.6\ mm$. (e) The variation of meniscus position with time for the experimental conditions shown in (a). (f) Experimental front-view images of the capillary flow meniscus at the step valve (SV) with the channel geometry and contact angle in (a).

The variation of meniscus position with time for the experimental conditions shown in Fig. 3a from the channel inlet ($x = 0$) is depicted in Fig. 3e. As shown in Fig. 3e and 2b, consistent with classical capillary flow theory, the meniscus initially exhibits an inertial regime characterized by a linear relationship with time, $x \sim t$. At later times, the flow transitions to the Lucas–Washburn regime, where the meniscus displacement scales with the square root of time, $x \sim \sqrt{t}$. Upon encountering the geometric step, the meniscus experiences an abrupt deceleration. In the expanded channel downstream of the step, the meniscus propagates at a reduced speed, attributed to the larger channel dimensions and consequently lower curvature. The dynamic contact angle formed by the advancing meniscus at the side walls exceeds the static contact angle of the liquid, following Tanner's law: $\theta_d = (\theta_s^3 + A\,Ca)^{\frac{1}{3}}$, where $A$ is a scaling constant and $Ca$ is the capillary number. For instance, under the experimental conditions shown in Fig. 3a, the dynamic contact angle is observed to be $45°$, corresponding to a static contact angle of $42°$. In both step valve (SV) and offset step valve (OSV) configurations under flow conditions, the meniscus undergoes complex shape transformations, and the associated contact line motion displays intricate dynamics. The top and front views of the meniscus motion obtained from experiments, are shown in Fig. 3(a-d) and 3f, respectively.

The isometric views of meniscus motion obtained from numerical simulations are presented in Fig. 4 (a-d). The top, side and front views of meniscus motion extracted from numerical simulations are presented in Fig. 4e and 4f. For the SV, from simulation results in Fig. 4a and 4e, it is observed that upon reaching the junction, the meniscus contact line remains pinned to the lower edge of the incoming channel but continues to advance along the top wall of the expanded section until the static contact angle ($\theta_s$) is reestablished at the upper boundary. In this pinned state, the contact angle formed with the bottom channel wall approximates $\frac{\pi}{2} + \theta_s$. During flow, the meniscus curvatures in the top and side planes, $R_{xy}$ and $R_{xz}$, progressively increase and shift from negative (concave) to positive (convex), while the curvature in the front plane, $R_{yz}$ remains negative. This curvature configuration generates a net positive Laplace pressure that drives the meniscus forward. The initial lateral (width-wise) motion of the contact line is attributed to corner flow, which occurs under the condition $\theta_s + \alpha < \frac{\pi}{2}$, where $\alpha$ represents half of the corner angle. In the present setup, $\alpha = 45°$, leading to the consistent observation of corner flow for $\theta_s < 45°$, as evidenced in Fig. 4a and 4e. Interestingly, corner flow persists for static contact angles as high as $\theta_s = 57°$ when the height of the incoming channel is smaller (see Fig. 4c).

A lower channel height corresponds to a smaller $R_{yz} \sim h_1$, resulting in a higher Laplace pressure, which may account for the sustained corner flow at elevated contact angles. For smaller step sizes in the width direction, $\Delta w = (w_2 - w_1)$, corner flow reaches the channel's side walls (see Fig. 4a and 4e). Similarly, for reduced step sizes in the height direction, the corner flow lines contact the side walls and propagate downward along the z-axis in the yz-plane at the junction, while also extending in the x-direction along the corner. The meniscus curvatures that contribute to the positive Laplace pressure driving the flow across the junction are inversely proportional to the step sizes in width and height, i.e., $R_{xy} \sim \Delta w$ and $R_{yz} \sim \Delta h$. Throughout the process, the meniscus remains concave (negatively curved), which maintains a high curvature and generates a positive Laplace pressure that facilitates meniscus progression across the xz-plane of the junction (see Fig. 4a and 4e). Conversely, when either the width or height step size ($\Delta w$ and $\Delta h$) is higher, the resulting decrease in curvature leads to a lower Laplace pressure. This reduction impedes the meniscus motion and results in only partial corner flow, thereby explaining the emergence of a no-flow regime (see Fig. 4b). Further, since the driving Laplace pressure at the meniscus is proportional to $cos\theta_s$ we find that corner flow can occur for a larger step size at a smaller contact angle. The above observations suggest that corner flow is a necessary condition for the onset of the flow regime.

As previously discussed, a no-flow regime typically occurs when there is a larger step size in either the width or height direction, or when the contact angle is high. Interestingly, our findings reveal that even with a significant step size, introducing an offset feature can induce a transition to a flow regime, as illustrated in Fig. 4c and 4f. In the offset-step Valve (OSV) configuration, one side wall of the incoming microchannel is extended relative to the other. This offset significantly alters the meniscus behavior, facilitating conditions that promote flow. Unlike the standard no-offset design, the OSV maintains a concave meniscus shape from the top view ($R_{xy}$) along with $R_{yz}$, exhibiting strong curvature that generates a higher positive Laplace pressure (see Fig. 4f). Notably, corner flow occurs even when the static contact angle $\theta_s$ exceeds 45°, indicating that the added offset contributes to the development of sufficient Laplace pressure. As a result, corner flow initiates along the shorter side of the microchannel at the top-left edge of the step, allowing the liquid to contact the side wall of the expanded channel. The meniscus then progresses along the side wall and downward, with curvature gradually increasing. Once a critical condition is reached, corner flow is also observed along the longer side wall, enabling the liquid front to contact the right wall. The meniscus then propagates through the expanded channel, establishing a flow regime. However, when the contact angle or step size becomes excessively large, even the OSV configuration fails to maintain continuous flow, leading instead to partial corner flow and the re-emergence of a no-flow regime, as seen in Fig. 4d.

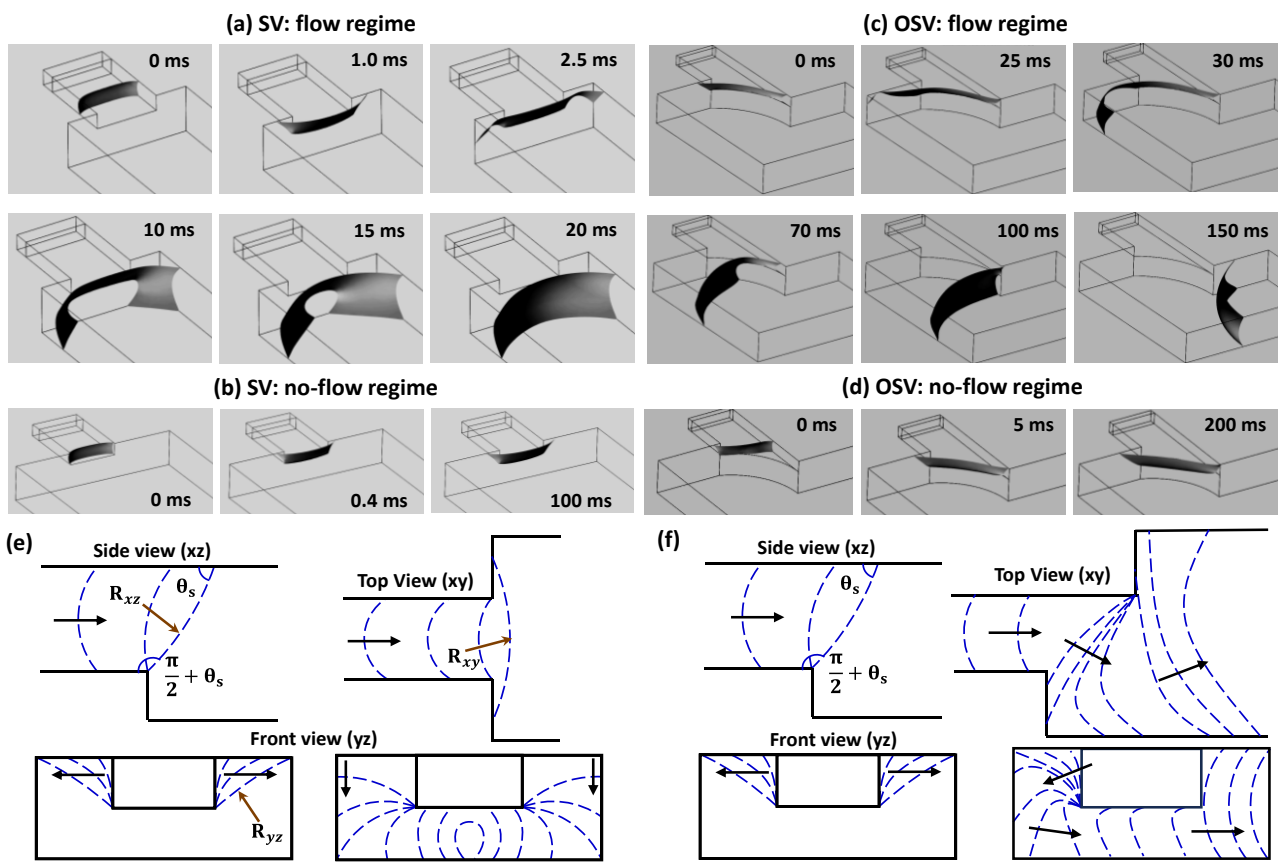

**Fig. 4:** Numerical simulation results showing the meniscus dynamics predicting the flow and no-flow regimes in case of stop valve (SV) and offset stop valve (OSV), (a) SV: isometric view, with $w_1 = 0.4\ mm$, $h_1 = 0.4\ mm$, $w_2 = 1.6\ mm$, and $h_2 = 0.5\ mm$, and $\theta_s = 42°$. (b) SV: isometric view, with $w_1 = 0.4\ mm$, $h_1 = 0.4\ mm$, $w_2 = 1.6\ mm$, $h_2 = 0.5\ mm$, and $\theta_s = 57°$. (c) OSV: isometric view with $w_1 = 0.4\ mm$, $h_1 = 0.3\ mm$, $w_2 = 1.6\ mm$, $h_2 = 0.5\ mm$, $\theta_s = 57°$ and $G = 0.6\ mm$. (d) OSV: isometric view with $w_1 = 0.4\ mm$, $h_1 = 0.3\ mm$, $w_2 = 1.6\ mm$, $h_2 = 0.5\ mm$, $\theta_s = 67°$ and $G = 0.6\ mm$. (e) SV: top, side and front views of the meniscus dynamics for the geometry in (a), (f) OSV: top, side and front views of the meniscus dynamics for the geometry in (c).

## B. Regime map: step valve (SV) and offset step valve (OSV)

The behavior of capillary flow at the step valve (SV), demonstrating both flow and no-flow conditions, is analyzed across various step dimensions in width and height, as well as contact angles, using a three-dimensional representation shown in Fig. 5a. The dimensionless parameters for step width and height are defined as: $\Delta w^* = (w_2 - w_1)/2w_1$ and $\Delta h^* = (h_2 - h_1)/h_1$. Our findings suggest that when the liquid contact angle ($\theta_s$) is less than 45°, a flow regime consistently occurs for smaller step ratios, specifically $\Delta w^* < 1$ and $\Delta h^* < 2$. Within this angle range, transitioning from flow to no-flow regimes occurs if the step width ratio exceeds 1.0 or the height ratio surpasses the critical value of 2.0. Interestingly, even when $\theta_s > 45°$, although the no-flow condition is predominant, flow regimes are still observed up to $\theta_s = 53°$ if the step dimensions are small. Conversely, for $\theta_s < 45°$, increasing step size in either direction can still lead to a no-flow regime. This implies that while a contact angle below 45° is necessary for flow, due to the onset of corner flow, step size also plays a critical role. Specifically, smaller steps enhance the meniscus curvature, promoting a favorable Laplace pressure that drives the fluid forward, thereby becoming a sufficient condition for flow.

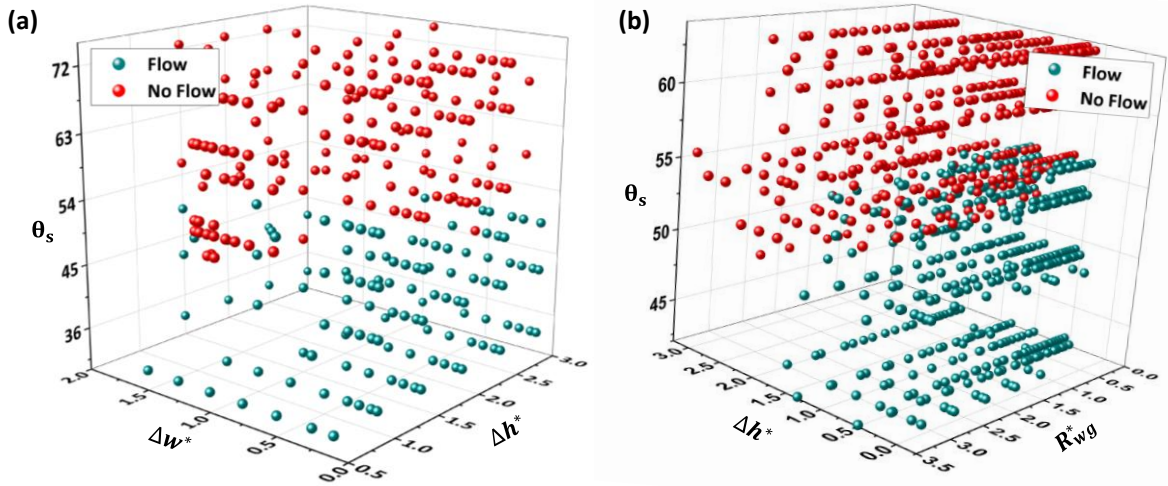

**Fig. 5:** Three-dimensional representation of flow and no-flow regimes in case of step valve (SV) and offset-step valve (OSV). (a) SV: in terms of dimensionless step width ($\Delta w^*$) and step height ($\Delta h^*$) and contact angle ($\theta_s$). (b) OSV: in terms of dimensionless offset parameter ($R_{wg}^*$) and step height ($\Delta h^*$) and contact angle ($\theta_s$). The ranges of dimensions: $h_1 = 0.1 - 0.4\ mm$, $w_1 = 0.1 - 0.4\ mm$, $h_2 = 0.2 - 0.8\ mm$, $w_2 = 0.6 - 1.6\ mm$, and $G = 0 - 0.8\ mm$. Contact angles: 36° to 72°.

For offset step valves (OSVs), the introduction of a geometric offset ensures that the meniscus curvature in the top view ($R_{xy}$) remains concave. When this curvature interacts with the channel sidewalls, along with the vertical curvature ($R_{yz}$), it generates a positive net Laplace pressure, thereby advancing the meniscus. The combined effect of the offset g and the channel width $w_1$ on meniscus curvature is captured through the off-set parameter $R_{wg} = \sqrt{w_1^2 + G^2}$, with its normalized form defined as $R_{wg}^* = R_{wg}/w_1$, as illustrated in Fig. 6b. Compared to standard step valves (SVs), OSVs demonstrate flow regimes even at contact angles as high as $\theta_s = 57°$, provided $R_{wg}^* > 2$. This confirms that the consistently concave meniscus in OSVs supports favourable corner flow. Furthermore, for a fixed contact angle, OSVs support flow regimes over a broader range of step heights and widths compared to SVs. Collectively, these results demonstrate that incorporating an offset into the step geometry significantly extends the operational envelope for capillary-driven flow, both in terms of permissible contact angles and geometric dimensions.

## C. Flow and no-flow regimes: meniscus pressure variation and energy scaling

The flow and no-flow regimes in both step valves (SV) and offset step valves (OSV) are elucidated based on the net area-averaged Laplace pressure difference, defined as $\Delta p = (p_{atm} - p_{liq})$, across the meniscus at various axial positions along the flow direction. These results, obtained from numerical simulations, are presented in Fig. 6a. For the SV configuration with $\Delta w^* = 1.5$, $\Delta h^* = 2.5$, and $\theta_s = 53°$, the Laplace pressure jump is initially positive within the incoming channel, indicative of capillary-driven flow. However, this pressure difference drops to zero at the step, consistent with the experimentally observed no-flow regime. In contrast, for another SV configuration with modified parameters ($\Delta w^* = 1$, $\Delta h^* = 2$, and $\theta_s = 42°$), a non-zero positive pressure jump is maintained at the step, correlating with the observed flow regime. Additionally, the pressure jump across the meniscus in the expanded channel is reduced compared to that in the incoming channel, which is attributed to an increase in meniscus radius of curvature in the wider channel section.

In the OSV case, with $\Delta h^* = 2$, and $\theta_s = 57°$, the flow remains arrested for offset values $R^*_{wg} = 0.5$, and 1.0 as the Laplace pressure jump at the step approaches zero. However, increasing the offset parameter to $R^*_{wg} = 2$ and 2.5, results in a positive and non-zero pressure jump at the step, enabling a transition to the flow regime, as shown in Fig. 6b. As the offset parameter $R^*_{wg}$ increases, the magnitude of the pressure jump at the step also increases. Notably, a secondary peak in the pressure profile (see Fig. 7) corresponds to the moment when the meniscus detaches from the side wall at the step and reattaches to the side wall of the expanded channel, enhancing the net driving pressure. Once the flow proceeds into the expanded channel, the pressure jump gradually declines and stabilizes, due to the larger and constant radius of curvature of the advancing meniscus.

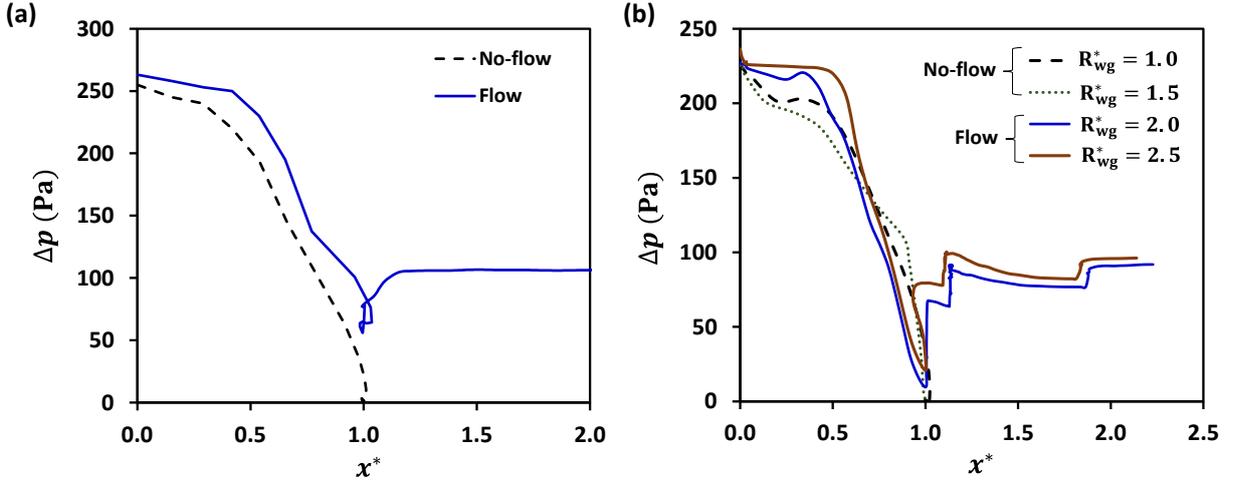

**Fig. 6:** Variation of Laplace pressure difference, $\Delta p = (p_{atm} - p_{liq})$, across the meniscus located at various axial positions along the flow direction, obtained from numerical simulations. (a) For the step valve (SV): no-flow regime with $\Delta w^* = 1.5$, $\Delta h^* = 2.5$, and $\theta_s = 53°$ and flow regime with $\Delta w^* = 1$, $\Delta h^* = 2$, and $\theta_s = 42°$. (b) For the offset-step valve (OSV) with $\Delta w^* = xx$, $\Delta h^* = 2$, and $\theta_s = 57°$: no-flow regime with $R^*_{wg} = 0.5$, and 1.0 and flow regime with $R^*_{wg} = 2.0$, and 2.5.

To further interpret these regimes, a theoretical energy-based model is developed to predict the transition between flow and no-flow states in both SV and OSV geometries. The analysis assumes that meniscus motion is governed by surface tension forces, with negligible contributions from inertial and viscous effects. This assumption is supported by the low values of the dimensionless parameters: the Weber number $We = \frac{\rho U^2 l}{\mu}$, and the Capillary number $Ca = \frac{\mu U}{\gamma}$, both of which satisfy $We < 0.01$ and $Ca < 0.02$, respectively [29]. The total surface energy of the system is calculated at two critical meniscus positions: (i) just before the step, where the meniscus is pinned at the edge of the incoming channel, and (ii) just after the meniscus crosses the step. These energies are used to determine the flow state based on the principle of minimum energy. A positive energy difference, $\Delta E = (E_2 - E_1) > 0$ indicates a no-flow regime, whereas a negative energy difference, $\Delta E < 0$ suggests flow.

At the first position, the total surface energy includes contributions from the solid-liquid, solid-gas, and liquid-gas interfaces:

$$E_1 = A_{sl-1}\,\gamma_{sl} + A_{sg-1}\gamma_{sg} + A_{lg-1}\gamma_{lg}$$

Here, $\gamma_{sl}$, $\gamma_{sg}$, and $\gamma_{lg}$ are the surface energies of the solid-liquid, solid-gas and liquid-gas interfaces, and $A_{sl-1} = 2(w_1 + h_1)\,L_1$, $A_{sg-1} = 2(w_2 + h_2)\,L_2$ and $A_{lg-1}$ are the surface areas of the solid-liquid, solid-gas and liquid-gas interfaces at the first location, where $w_1$ and $h_1$ and $L_1$ are the width, height and length of the incoming microchannel, $w_2$, $h_2$ and $L_2$ are the width, height and length of the expanded microchannel after the step. The liquid-gas surface area is approximated as, $A_{lg-1} = R_{11}R_{12}\theta_s^2$, where $R_{11} = \frac{h_1}{2cos\theta_s}$, and $R_{12} = \frac{w_1}{2cos\theta_s}$, are the radii of curvature of the meniscus at the first position in the height and width directions. Here $\theta_s$ is the static contact angle made by the liquid with the channel walls.

Similarly, at the second location, the total surface energy of the meniscus can be obtained. The surface areas of the solid-liquid, solid-gas and liquid-gas interfaces at the second position, after crossing the step are expressed as: $A_{sl-2} = 2(w_1 + h_1)\,L_1 + 2(w_2 + h_2)\,\delta l$, $A_{sg-2} = 2(w_2 + h_2)\,(L_2 - \delta l)$ and $A_{lg-2} = R_{21}R_{22}\theta_s^2$, where $R_{21} = \frac{h_2}{2cos\theta_s}$, and $R_{22} = \frac{w_2}{2cos\theta_s}$, are the radii of curvature of the meniscus at the second position in the height and width directions.

From the above, the difference in the surface energies of the meniscus at the first and the second locations is expressed as,

$$\Delta E = \frac{\theta_s^2}{4cos^2\theta_s}(w_2 h_2 - w_1 h_1) - 2\delta l\,(w_2 + h_2)\,\gamma_{lg}cos\theta_s$$

The above expression is applicable to both the step-valve (SV) and off-set step valve (OSV) cases. However, the radius of curvature $R_{12}$ in the OSV case is expressed as, $R_{12} = \frac{\sqrt{w_1{}^2 + G^2}}{2cos\theta_s}$, and therefore

$$\Delta E = \frac{\theta_s^2}{4cos^2\theta_s}\left(w_2 h_2 - \sqrt{w_1{}^2 + G^2}\,h_1\right) - 2\delta l\,(w_2 + h_2)\,\gamma_{lg}cos\theta_s$$

This expression simplifies to the SV case when $G = 0$. For channels with constant cross-section, $w_1 = w_2$ and $h_1 = h_2$, $\Delta E < 0$ always holds, indicating spontaneous capillary flow for $\theta_s < 90°$. However, as the cross-sectional area of the expanded channel increases significantly with $w_2 h_2 \gg w_1 h_1$, $\Delta E$ becomes positive, favouring a no-flow regime. Importantly, introducing a non-zero offset ($G > 0$) can reverse this effect, rendering $\Delta E$ negative and restoring flow.

Experimental validation for the scaled model for the SV case is provided through a 2D regime map, as shown in Fig. 7a, and Supplementary Material, illustrating the transition between flow and no-flow regimes as a function of step width ($\Delta w^*$) and height ($\Delta h^*$) for varying contact angles $\theta_s$. The transition boundary aligns well with the $\Delta E = 0$ line predicted by the model, with flow observed when $\Delta E < 0$ and no-flow when $\Delta E > 0$. Moreover, as $\theta_s$ increases, the $\Delta E = 0$ boundary shifts toward smaller step sizes, confirming that higher contact angles reduce the tolerance for geometrical discontinuities, consistent with experimental observations. The experimental validation of the model for the OSV configuration is depicted in Fig. 7b, which also shows a good agreement. Thus, the energetic criterion captures the influence of geometrical parameters and liquid contact angle to successfully predict the experimentally observed flow transition boundaries in both SV and OSV cases.

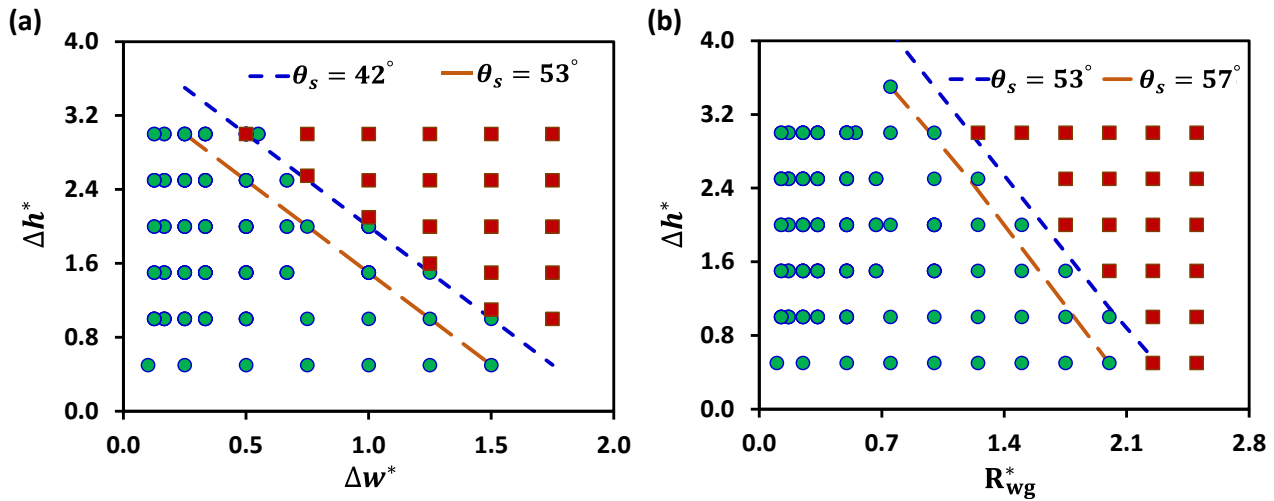

**Fig. 7:** Experimental validation of the scaled model for predicting flow and no-flow regimes. The symbols indicate experimental points, while the $\Delta E = 0$ line from the scaled model marks the boundary between flow and no-flow regimes. (a) SV: step width ($\Delta w^*$) and step height ($\Delta h^*$), $\theta_s = 42°$, and $53°$. (b) OSV: offset parameter ($R_{wg}^*$) and step height ($\Delta h^*$), $\theta_s = 42°$, and $53°$.

## D. Flow synchronization in parallel microchannels

Achieving synchronization of capillary-driven flows at the microscale continues to remain a challenge [30]. Here, we present a combined implementation of step valves (SV) and offset-step valves (OSV) to realize synchronized capillary flow across a network of parallel microchannels [31]. Specifically, we demonstrate how introducing OSVs into the system architecture mitigates void formation caused by asynchronous meniscus propagation. The meniscus dynamics within a network of 7 parallel microchannels (numbered from top to bottom), with $w_1 = 0.4\ mm$, $h_1 = 0.4\ mm$, and $h_2 = 0.5\ mm$, and $\theta_s = 42°$, each equipped with step valves configured to operate in the flow regime at the downstream ends, is depicted in Fig. 8a (see Movie S5). Due to fabrication-induced variations in surface properties, the liquid menisci propagate at different velocities across the channels. Notably, in channel number $2 - 7$, the meniscus exits the step valve and contacts the outlet plenum wall prior to the arrival of the meniscus at the step valve in channel number 1. This temporal mismatch traps air ahead of the step, in the channel 1, leading to stagnation of the meniscus and resulting in void formation, ultimately disrupting synchronized capillary flow within the microchannel network.

To address this, we implement an offset-step valve, with $w_1 = 0.4\ mm$, $h_1 = 0.4\ mm$, $h_2 = 0.5\ mm$, $\theta_s = 57°$ and $G = 0.6\ mm$, at the end of the channel 1 (configured for flow), while retaining conventional step valves, with $w_1 = 0.4\ mm$, $h_1 = 0.4\ mm$, $h_2 = 0.5\ mm$, and $\theta_s = 57°$, configured for no-flow, at the remaining channels, i.e., channel number $2 - 7$. The corresponding meniscus dynamics are presented in Fig. 8b (see Movie S6). As expected, inhomogeneity in surface characteristics continues to produce variation in meniscus velocity. However, in the SV-equipped channels, the menisci halt at the step as designed. In contrast, the meniscus in the OSV-equipped channel advances past the step into the outlet plenum. To promote synchronized flow activation across all channels, a phase guide [32] is incorporated into the outlet plenum. This feature directs the advancing meniscus from the OSV-equipped channel toward the step valves of neighbouring channels. Upon contact, the halted menisci are reactivated, triggering simultaneous advancement and establishing synchronization across the entire microchannel array. Therefore, a combination of step valve (SV) and off-set step valve alongside a phase guide can be effectively implemented for the synchronization of capillary flow in parallel microchannels.

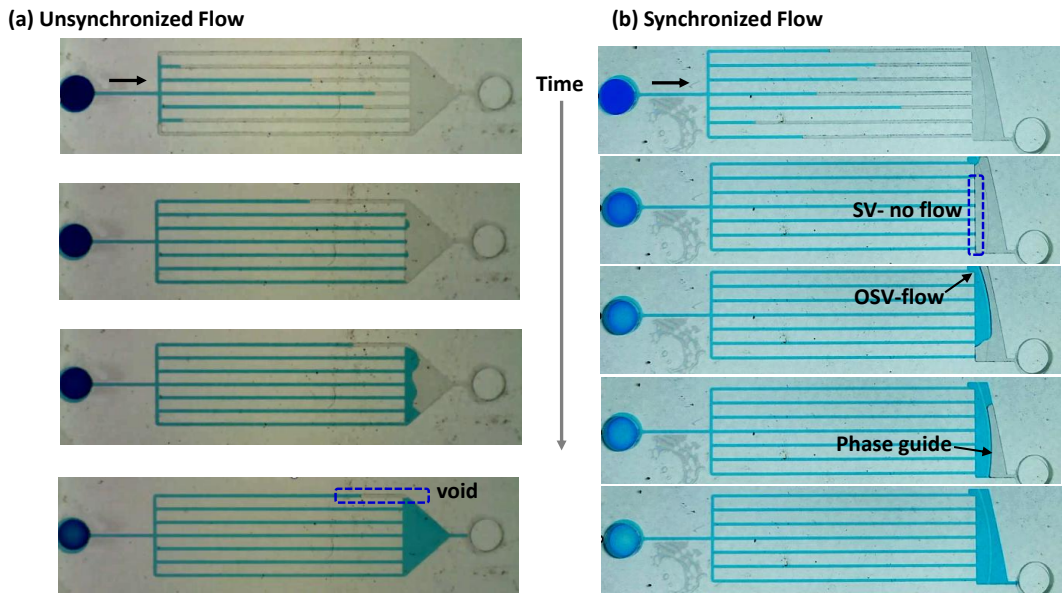

**Fig. 8:** Experimental images showing (a) Unsynchronized flow within a network of 7 parallel microchannels, each equipped with step valves (SV) configured to operate in the flow regime, $w_1 = 0.4\ mm$, $h_1 = 0.4\ mm$, and $h_2 = 0.5\ mm$, and $\theta_s = 42°$. (b) Synchronized flow achieved by implementing an offset-step valve (OSV) (configured for flow), while retaining conventional SV, configured for no-flow, in the other channels. SV: $w_1 = 0.4\ mm$, $h_1 = 0.4\ mm$, $h_2 = 0.5\ mm$, and $\theta_s = 57°$. OSV: $w_1 = 0.4\ mm$, $h_1 = 0.4\ mm$, $h_2 = 0.5\ mm$, $\theta_s = 57°$ and $G = 0.6\ mm$.

## IV. CONCLUSION

We have presented a comprehensive experimental and theoretical investigation of capillary flow in rectangular microchannels encountering geometric steps, revealing how wettability and step geometry jointly dictate flow stability. By systematically mapping the flow and no-flow regimes, we show that for contact angles below 45°, meniscus advancement occurs through corner flow consistent with the Concus–Finn condition, and that flow is favoured at smaller normalized step ratios ($\Delta w^* < 1$, $\Delta h^* < 0.5$). Remarkably, even at higher contact angles (up to ~53°), flow can persist when step dimensions are sufficiently small to maintain a negative meniscus curvature that generates a favourable Laplace pressure. Introducing offset step valves (OSVs) significantly extends the accessible flow regime. The geometric offset preserves a concave meniscus across the discontinuity, sustaining a positive net Laplace pressure and enabling flow even at $\theta_s = 57°$. An energy-based model predicts this transition quantitatively, showing that flow occurs when the surface energy decreases across the step ($\Delta E < 0$), in excellent agreement with experiments and simulations. Finally, we demonstrate synchronized capillary transport in parallel microchannel networks by integrating SVs and OSVs. The offset geometry mitigates velocity mismatches and enables self-triggered reactivation of pinned menisci, achieving robust, passive synchronization without external control. Beyond its direct utility for lab-on-a-chip and lateral flow diagnostic platforms, this work establishes a predictive energetic framework for geometry-mediated capillary stability in confined flows. The ability to tune Laplace-driven transitions through designed asymmetry offers new routes for developing self-regulating, autonomous microfluidic systems that bridge the gap between fundamental interfacial physics and practical microdevice engineering.

## ACKNOWLEDGEMENTS

The authors acknowledge support from IIT Madras via project no. RF21220988MERFIR008509, and by the Ministry of Human Resources and Development, Government of India, through the Institute of Eminence (IoE) project no. SB22231233MEETWO008509 via grant no. 11/9/2019-U.3(A).